\documentclass[twocolumn,amssymb,nobibnotes,aps,prl,reprint]{revtex4-1}

\usepackage[separate-uncertainty=true,multi-part-units=single]{siunitx}
\usepackage{graphicx}
\usepackage{bm}
\usepackage[usenames,dvipsnames]{color} 

\usepackage{array}
\newcolumntype{L}[1]{>{\raggedright\let\newline\\\arraybackslash\hspace{0pt}}m{#1}}
\newcolumntype{C}[1]{>{\centering\let\newline\\\arraybackslash\hspace{0pt}}m{#1}}
\newcolumntype{R}[1]{>{\raggedleft\let\newline\\\arraybackslash\hspace{0pt}}m{#1}}

\usepackage{cmap} 
\usepackage[T1]{fontenc}

\setcitestyle{round}
\makeatletter
\renewcommand\NAT@biblabelnum[1]{(#1)}
\makeatother

\begin{document} 

\title{QuipuNet: convolutional neural network for single-molecule nanopore sensing}

\author{Karolis Misiunas}
\email{karolis@misiunas.com}
\author{Niklas Ermann}
\author{Ulrich F. Keyser}
\email{ufk20@cam.ac.uk}
\affiliation{Cavendish Laboratory, University of Cambridge, UK}
\date{\today}

\begin{abstract} 
    Nanopore sensing is a versatile technique for the analysis of molecules on the single-molecule level. However, extracting information from data with established algorithms usually requires time-consuming checks by an experienced researcher due to inherent variability of solid-state nanopores. Here, we develop a convolutional neural network (CNN) for the fully automated extraction of information from the time-series signals obtained by nanopore sensors. In our demonstration, we use a previously published dataset on multiplexed single-molecule protein sensing~\cite{Bell2016a}. The neural network learns to classify translocation events with greater accuracy than previously possible, while also increasing the number of analysable events by a factor of five. Our results demonstrate that deep learning can achieve significant improvements in single molecule nanopore detection with potential applications in rapid diagnostics.
\end{abstract}

\maketitle

\subsection{ Introduction }

\begin{figure*}[htb!]
    \centering
    \includegraphics[width=\linewidth]{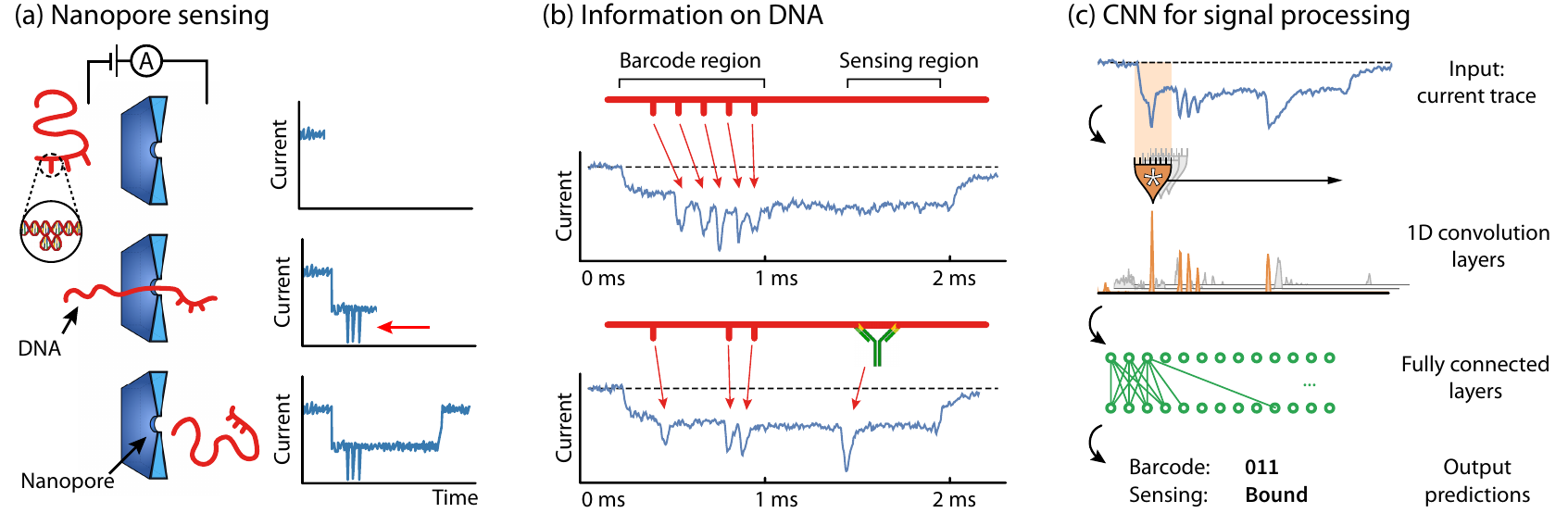}
    \caption{\label{fig_introduction}
    Convolutional neural networks for the analysis of nanopore data.
    (a) The shape of molecules is contained in the time-dependent ionic current signal from passing the molecules through a nanopore.
    For example, a molecule with three protrusions passing through the nanopore leads to a current event with three secondary current drops as indicated by the red arrow. 
    (b) Current traces associated with the modified DNA molecules from~\cite{Bell2016a}. The first half of the molecule encodes a unique barcode: the first peak marks the start; three bits uniquely identify molecule design; and the last peak signifies the end. The two events have barcodes `111' and `001'. The second half has a binding site for a specific molecular target.
    (c) Data analysis using deep learning methods: convolution layers extract local features such as current drops of different width (shown in orange). The features are interpreted by a fully connected neural network, which outputs a prediction for the barcode and the target binding state. 
    } 
\end{figure*}

Nanopores have emerged as powerful sensing devices for single molecules~\cite{Shi2017,Albrecht2017a}, with applications in DNA sequencing~\cite{Loman2015}, protein detection~\cite{Bell2016a,Wei2012a,Kong2016a,Celaya2017,Morin2014patent,Sze2017}, the study of protein folding~\cite{Si2017a}, 
SNP genotyping~\cite{Kong2017}, 
data storage~\cite{Morin2014patent}, and DNA computing~\cite{Ohara2017}. A typical setup consists of two liquid filled reservoirs connected by a nanopore with diameters down to a few nanometres. An external electric field drives charged molecules through the nanopore, as shown in Figure~\ref{fig_introduction}a. The passage of molecules modulates the current, producing a characteristic signal that contains information about the shape of the molecule.

The readout is a time-series current trace corresponding to the shape of the molecule, usually called an event. Detection of such events can be achieved using simple current thresholds, but the subsequent analysis of features within each identified event is often made difficult by a poor signal-to-noise ratio, varying conformation of the molecule, and non-specific interactions with the nanopore surface. For example, Figure~\ref{fig_introduction}b shows two events from a multiplexed protein sensing technique published in~\cite{Bell2016a}. 
The authors used a DNA molecule as a carrier for a protein target. Modifications along the DNA molecule and bound targets produce secondary current drops during the translocation event, as shown in the two traces. In the first half of the structure, DNA hairpin loops at defined positions and their corresponding secondary drops were used to encode a digital barcode. This barcode uniquely identified a binding site in the other half of the DNA molecule. The presence of a target at the binding site could be inferred from a single secondary drop in the second half. This approach allows the simultaneous detection of a large number of targets, only limited by the number of distinct barcodes. The information is encoded in additional current drops during the event, much like the knots on a string used in the Inca Quipu system~\cite{ascher2013mathematics}.

Analysis of the event data requires accurate detection and subsequent interpretation of secondary current drops~\cite{Bell2016a}. However, simple peak finding algorithms often fail at reliably classifying large parts of the data. Common causes of errors are a varying peak magnitude, noise~\cite{Kong2016a}, fluctuating velocities~\cite{Bell2016b}, overlapping peaks, DNA knots~\cite{Plesa2016} and folded molecules. To mitigate these effects the nanopore community has developed sophisticated algorithms~\cite{Henley2016,Forstater2016a,Raillon2012,Plesa2015}. However, they frequently require manual parameter tuning for each dataset and supervision of algorithms~\cite{Bell2016a,Sze2017}. In the worst case scenario, researchers have to manually interpret the data, leading to small sample sizes, possible confirmation bias, or data analysis duration exceeding measurement time.

In this paper, we show that deep learning is ideally suited for automating the analysis of nanopore sensing data. For our study, we use the previously mentioned multiplexed protein sensing dataset from~\cite{Bell2016a}. The dataset contains separate control measurements for each specific barcode, without other bit permutations present in the solution. This automatically provides labelled data to train the supervised learning model. At the same time, the data is sufficiently complex to require an elaborate algorithm for the classification of events. In~\cite{Bell2016a} a twelve step approach was used to identify the bit sequence and presence or absence of a target on each DNA construct. That method relied on more than a dozen manually adjusted parameters that were carefully optimised, but still it could only use a small fraction ($\sim 20\%$) of events, discarding up to 80\% of the difficult-to-interpret events that failed some predefined set of criteria. 
Here, we show that machine learning models are able to interpret and classify data without the need for manual tuning and the development of complex algorithms while increasing the number of usable events by a factor of five. Our implementation is open-source and available online to enable the adaptation of deep learning to other nanopore sensing problems~\cite{QuipuNet2018}.

\subsection{ Methods }

We chose convolutional neural networks (CNN) as the machine learning approach because of their suitability for detecting local patterns~\cite{LeCun1998,Goodfellow-et-al-2016}. A recent study showed that CNNs perform well on simulated current traces from an STM tunnel junction~\cite{Albrecht2017}. For comparison,
DNA bases can be accurately determined from current levels using recurrent neural networks~\cite{Boza2017}. However, our goal is fundamentally different, as we are trying to identify the pattern encoded on the DNA secondary structure from a  variable nanopore system. Therefore, we chose to use the CNN architecture. 
A typical CNN consists of two parts, as shown in Figure~\ref{fig_introduction}c. First, a series of convolutions are applied to the raw input data. Then a dense neural network learns to interpret the processed signal. 
The output is a prediction about which class a particular input belongs to. In our case, the prediction is a barcode on the DNA constructs and whether a target has bound to it.

Before feeding the data into the neural network, we perform two preparation steps. First, the raw dataset contains erroneous detections, caused by contaminations, incomplete DNA fragments, and non-specific interactions with the pore walls. We use standard filtration methods to remove these detections~\cite{Mihovilovic2013}: we exclude events whose area under the current trace (electronic charge deficit) lies outside two standard deviations of the mean, as well as those with current drops larger than $3.2\times$ the unfolded event current level. Details are available in the supplementary material~\cite{QuipuNet2018}. This filtration removes up to 30\% of the detections recorded with the measurement setup. 
After filtration we still observe some events with errors, such as a missing bit in the barcode structure. Therefore, perfect accuracy is unattainable using realistic datasets.

Secondly, we want the model to identify a molecule, but not the experiment. The problem arises because nanopores vary in shape and conductivity, leading to a  correlation between events measured with the same nanopore. It is possible for the neural network to over-fit to these variations, thereby learning to identify a nanopore instead of the barcode on a molecule. To reduce such over-fitting, we normalise the events from each nanopore to have the same unfolded current level (arbitrarily set to $-1$). In addition, we test the model using independent experiments to reduce the chances of spurious correlations. Table~\ref{table_data} shows the number of events in the training and test sets.

\begin{table}[b!]
    \centering
    \sisetup{table-format=-1.6,
    separate-uncertainty=false
    }
    \begin{tabular}{l|R{1.1cm}  R{1.1cm}  R{1.3cm} R{1.3cm}}
        \multicolumn{1}{c}{} & \multicolumn{2}{c}{Event No.}  &  \multicolumn{2}{c}{Experiment No.} \\ 
        \hline \hline
        Label            & Train & Test & \shortstack{without \\ protein} & \shortstack{with \\ protein}\\
        \hline
        000              & 5593  & 253  & 5  & 0 \\
        001              & 8155  & 502  & 3  & 4 \\
        010              & 2319  & 101  & 4  & 0 \\
        011              & 15178 & 827  & 4  & 7 \\
        100              & 876   & 83   & 3  & 0 \\
        101              & 7251  & 427  & 2  & 4 \\
        110              & 6473  & 606  & 5  & 0 \\
        111              & 6680  & 665  & 5  & 2 \\
        \hline
        Unbound          & 36551 & 2191 & 31 & 0 \\
        Bound            & 15874 & 1273 & 0  & 17 \\
        \hline \hline
        Total            & 52525 & 3464 & 31 & 17\\
    \end{tabular}
    \caption{
    The number of events in the training and testing sets. The last two columns show the number of independent experiments without protein (unbound state) and with protein (bound state).  
    }
    \label{table_data}
\end{table}

\begin{figure}[bh!]
    \centering
    \includegraphics[width=8cm]{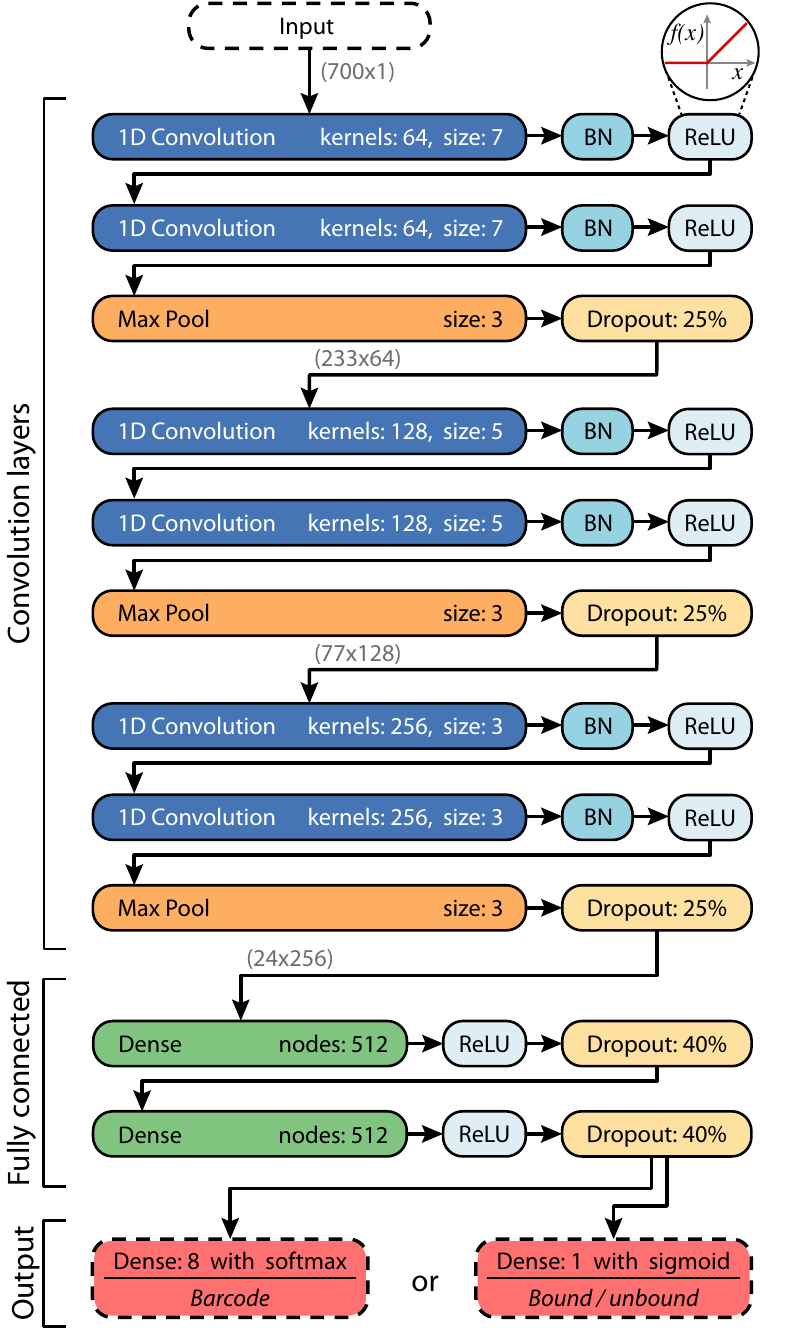}
    \caption{\label{fig_model}    
    The architecture of the neural network, where each element is briefly described in the methods section. 
    Acronyms:  
    BN~is a batch normalisation layer; ReLU is a rectified linear unit and is shown on top. 
    Numbers in the brackets correspond to the matrix sizes encoding a single event at that point in the network. The model has \si{3\,995\,920} trainable parameters.
    } 
\end{figure}

As mentioned above, our predictor model is based on a machine learning technique called neural networks. The architecture of such a network specifies how the network nodes are connected and what operations are applied. In order to find a suitable architecture for nanopore data, we investigated different alternatives by educated trial and error. The model presented here is inspired by the image classification network in~\cite{VGG2014}, which we modified to perform 1D convolutions. Figure~\ref{fig_model} shows the architecture. 

We optimised the (hyper-)parameters to work well for nanopore data by trial and error. A typical procedure is to pick one hyper-parameter, such as the number of convolution layers, then increase the number and measure the resulting accuracy. If the accuracy increases we stick with the new number, but if it decreases or does not change we stick with the old number. We then pick a different hyper-parameter and repeat the procedure. To avoid over-fitting to the test, we measured the accuracy gains using a development set, which is independent from the test set and 20 times smaller than the training set. The reported numbers in Figure~\ref{fig_model} are the result of our optimisation.

The input for the neural network is a current trace from a measurement event. 
The data from~\cite{Bell2016a} produced events with an average length of $402$ data points. This includes short stretches of current recording before and after the event. 
As the maximum length of the event never exceeded $700$ points, 
we use a $700$-element vector as the input. The shorter events are padded at the end with Gaussian noise ($\mu=0$, $\sigma=0.072$, corresponding to average noise levels).

Each box in Figure~\ref{fig_model} corresponds to a so-called `hidden layer' that performs a specific task and passes on the information. Here, we give a brief description of each component; we refer interested readers to the machine learning literature for more details~\cite{LeCun1998,Goodfellow-et-al-2016}.

Convolution layers extract features with local structure, such as peaks or steps. These layers perform a discrete convolution on a segment of the input by multiplying it with a small window, called kernel, and moving along to the next segment (stepping by a single vector element). The output is large if the input features match the kernel, where its weights are learned from the training data. For example, Figure~\ref{fig_introduction}c shows the output after the first convolution layer, where the orange line corresponds to a kernel that detects peaks. Other kernels detect other features in the input data, which are often difficult to interpret, as seen by two grey lines that correspond to different kernels.
After each convolution, we apply a batch normalisation (BN) layer that normalises the data to have zero mean and unit variance~\cite{Ioffe2015}. These layers improve our network training convergence. Finally, an activation function is applied -- a piecewise function called rectified linear unit (ReLU), $f(x) = \max(0, x)$. 
This non-linear function is necessary for learning non-linear relationships between features~\cite{Goodfellow-et-al-2016}. The activation function completes one row in the diagram, its output goes into the next convolution layer.

Roughly speaking, the deeper layers capture more abstract and complex features. We follow the common practice of increasing the number of kernels for deeper layers~\cite{LeCun1998}: from 64 to 128, then to 256. Each step doubles the amount of information passed to the next layer. For every two convolutions, we have a `Max Pool' layer to reduce the amount of information by down-sampling spacial dimensions. A Max Pool layer splits an input vector into segments of three numbers and returns only the maximum values within the segment. This arrangement is believed to improve spacial invariance for feature extraction~\cite{Goodfellow-et-al-2016}.

The dropout layer reduces over-fitting by randomly switching off a fraction of nodes in the layer above. This encourages the network to learn more robust features that do not depend on a single node~\cite{Hinton2012}. Note that the dropout is only applied during training, because we want maximum accuracy while using the algorithm.

The second half of the network is a densely connected neural network with two hidden layers and a ReLU activation function. In a dense network, the nodes between adjacent layers are fully connected, as illustrated in Figure~\ref{fig_introduction}c. The weights for these connections are learned from the training data.

The output layer is adjusted depending on the task. In our case, we have two outputs: the barcode and sensing region. The barcode output is a vector with 8 elements and a softmax activation function.  The softmax normalises the output vector to have a sum of one such that each element is a proxy for the probability for a different barcode. We take the maximal value to be the predicted barcode. 
For the sensing region, the output is a single number with a sigmoid activation function. This number is a proxy for the probability of having a target bound to the sensing region. Note that these are two networks that are trained separately and give independent outputs.

The model is trained for 200 epochs on a GPU (Nvidia GeForce GTX 1080 TI). The aim of the training is to find the weights that maximise accuracy, which corresponds to minimising a loss function. For barcodes, the loss function is categorical cross-entropy, while for the sensing region it is binary cross-entropy. To minimise the loss function we use the Adam optimisation algorithm~\cite{Kingma2014} (LR=$0.001$; decay=$0.97$; batch size of $32$). 
Typical training takes \SI{200}{\min}, while evaluation is much faster at \SI{1600}{\text{events}\per\second}, making QuipuNet suitable for real-time classification.

\subsection{ Results }

\begin{figure}[htb!]
    \centering
    \includegraphics[width=\linewidth]{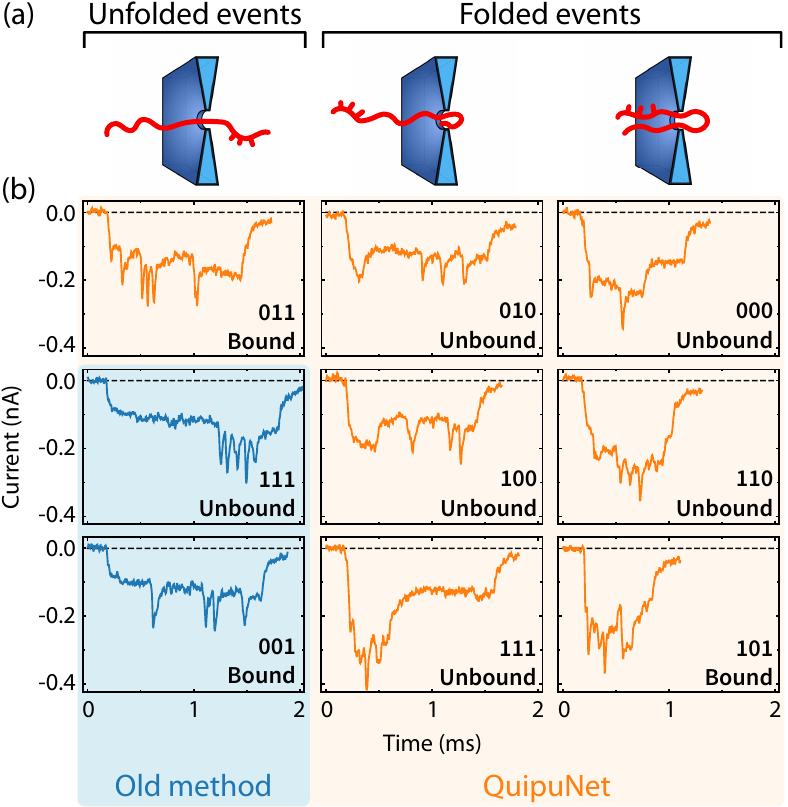}
    \caption{\label{fig_problem}
    Example events identified by semi-automated algorithm~\cite{Bell2016a} and QuipuNet. The sketches in (a) show some of the possible DNA configurations during the passage through a nanopore. The shape of the molecule complicates the semi-automated analysis~\cite{Bell2016a}. 
    (b) These 9 example events present typical results from the data set in~\cite{Bell2016a}. The original algorithm only identified the two blue events: 111, unbound and 001 bound; while QuipuNet correctly identified all these events. It is important to note here that QuipuNet increased the number of usable events by a factor of $\sim 5$.
    } 
\end{figure}

QuipuNet correctly identifies almost all events even with highly complex shapes, as shown in Figure~\ref{fig_problem}. For example, the first event in column one enters the nanopore with the barcode first, while the second and third examples enter with the sensing region first. QuipuNet can interpret both directions. Columns two and three show that it learns to identify folded DNA events which occur when a nanopore captures the DNA molecule somewhere along its length. These events are particularly difficult to interpret because there are many possible outcomes and peaks tend to be less pronounced. For comparison, the method from~\cite{Bell2016a} discarded folded events so that only the events shown in blue could be identified.

\begin{table}[ht!]
    \centering
    \sisetup{table-format=-1.6,
    separate-uncertainty=false
    }
    \begin{tabular}{l*{6}{r}}
                    & Precision & Recall & \shortstack{Data \\ utilised} \\
        \hline
        \bf{Barcode readout} \\
        \hline \hline
        Bell \& Keyser~\cite{Bell2016a} & 0.937 & 0.182 & 0.194  \\
        Human                           & 0.978 & 0.440 & 0.450   \\
        QuipuNet (all data)             & 0.946 & 0.946 & 1.000  \\
        QuipuNet (best 80\%)            & 0.987 & 0.789 & 0.800   \\

        \hline
        \bf{Sensing region} \\
        \hline \hline
        Bell \& Keyser~\cite{Bell2016a} & 0.940 & 0.192 & 0.204   \\
        Human                           & 0.931 & 0.405 & 0.435   \\
        QuipuNet (all data)             & 0.971 & 0.971 & 1.000  \\
        QuipuNet (best 80\%)            & 0.997 & 0.798 & 0.800   \\
    \end{tabular}
    \caption{
    Performance comparison between QuipuNet and other methods. 
    Precision is the fraction of correctly identified samples out of attempted guesses while recall gives the fraction of correctly identified samples out of all the events. 
    Data utilised is a fraction of events that the algorithm attempted to identify. 
    }
    \label{table_results}
\end{table}

\begin{figure}[htb!] 
    \centering
    \includegraphics[width=\linewidth]{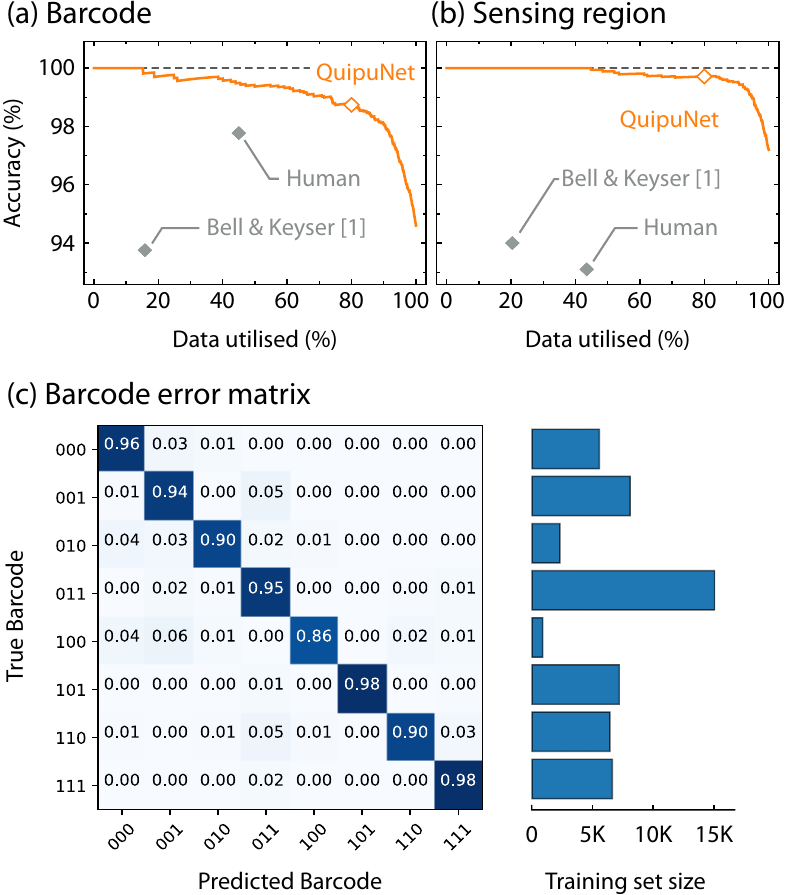}
    \caption{\label{fig_results}
    Evaluating the performance of QuipuNet.
    (a) Barcode prediction accuracy (precision) as a function of data utilised. The accuracy increases when the least confident predictions are removed. 
    (b) Sensing region prediction accuracy as a function of data utilised. 
    (c) Error matrix: rows represent true barcodes from the test set, while columns are the barcodes that QuipuNet assigned them to. In an ideal case, it would be a diagonal matrix. The matrix was evaluated using the entire test set.
    On the right, bars show the number of events in the training set for each barcode. Accuracy correlates with the size of the training set. 
    } 
\end{figure} 

Table~\ref{table_results} presents a quantitative comparison of accuracy. The first metric for accuracy is \emph{precision}, which gives the fraction of correctly identified events out of attempted guesses. Precision can be boosted by refusing to label difficult events. On the other hand, the \emph{recall} metric gives the fraction of correctly identified events out of all the events (after filtration). For example, the Bell \& Keyser method~\cite{Bell2016a} and human experts achieve high accuracy but have a low recall because events with ambiguous barcode patterns are discarded.

QuipuNet achieves a precision of $0.946$ for barcodes and $0.971$ for the sensing regions. This is $1.0\%$ and $3.4\%$ higher than the Bell \& Keyser method.
A much bigger difference can be seen in the recall metric because QuipuNet classifies all the data. The recall is five times larger than the original method~\cite{Bell2016a} for both the barcode and sensing region. These results suggest that QuipuNet accurately classifies the nanopore event data, including folded events. As a result, QuipuNet outputs 5 times more data than the previous method for the same experiments.

To measure human expert performance, one of the authors labelled 500 randomly chosen events and compared them with the true labels (it took around one hour). Only 45\% of events could be labelled reliably because of the ambiguity introduced by folds or overlapping peaks. Compared with human performance, QuipuNet is $3.3\%$ less precise at reading the barcode and $4.4\%$ better at reading the sensing region. In both cases, the recall metric is more than twice that of a human expert. 

To optimise for accuracy, we can discard low confidence predictions to increase the precision. Practically, it makes sense to discard events where a barcode is simply missing or otherwise impossible to identify. To achieve this, we estimate the confidence using the maximal value of the softmax output vector and then discard events with the lowest confidence. We use a `data utilised' fraction to show how much data remains after discarding low confidence predictions.

Figure~\ref{fig_results}a shows the accuracy as a function of data utilised for the barcode predictions (evaluated on the test set). The accuracy increases with the amount of discarded data, suggesting that the confidence estimator correctly identifies poor predictions. The accuracy curve is significantly above manual labelling and the Bell \& Keyser method, suggesting that QuipuNet outperforms both. For illustrative purposes, at 80\% utilised data QuipuNet precision is $0.987$, which is higher than the human performance. Figure~\ref{fig_results}b shows an equivalent plot for the sensing region predictions. Here, QuipuNet achieves a nearly perfect precision of $0.997$ for 80\%  utilised data. In both cases, discarding low confidence predictions increases the accuracy of the QuipuNet algorithm.

The predictions for the sensing region have a higher accuracy than those for the barcodes. We attribute this to two effects. First, the sensing region typically has a higher signal-to-noise ratio, i.e. larger current drops. Secondly, the barcode prediction is an intrinsically harder problem, because the algorithm must distinguish between 8 different classes, instead of two.

Figure~\ref{fig_results}c shows where the errors are made for the barcode predictions. The matrix suggests that QuipuNet makes more mistakes for certain barcodes. For example, the prediction for barcode `100' has a precision of only $0.86$, which can be attributed to the small training set. It only has 876 events measured by two experiments while the third experiment was used for the test set. A larger training set is expected to improve the accuracy.

The error matrix also provides insights for designing more robust barcodes. The barcodes `000', `001', `101', `110', and `111' all have a similar amount of training data, but the symmetric barcodes have a higher accuracy. Here, symmetric barcodes are `000', `101', and `111' (`010' has smaller amount of training data). This observation suggests that using only symmetric patterns for barcodes might improve the overall accuracy.

\begin{figure}[htb!] 
    \centering
    \includegraphics[width=\linewidth]{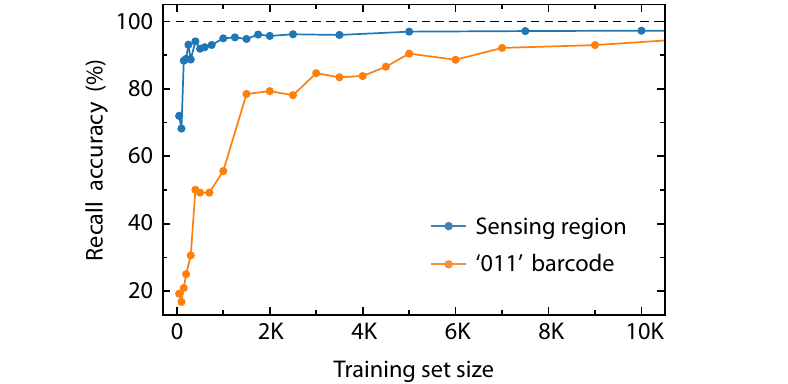}
    \caption{\label{fig_data_quantity}
    Recall accuracy as a function of training set size. The number of events are shown for bound/unbound states and the `011' barcode.  
    } 
\end{figure}

Finally, we trained QuipuNet on a reduced training set to assess the relationship between accuracy and training set size, as shown in Figure~\ref{fig_data_quantity}. 
For the sensing region, we randomly picked the same number of events for bound and unbound states. For the barcode, we randomly reduced the training set size of the `011' barcode to a number specified on the $\hat x$ axis, while the other barcodes had the same number of events as specified in Table~1. 
The resulting recall metric reaches $80\%$ at $2000$ training events, $90\%$ at $8000$ and then slowly increases to $>90\%$ for more than $8000$ training events. The increase in accuracy beyond $90\%$ appears to be asymptotic and would require even 
larger training sets.
The classification of the sensing region (blue data in Figure~\ref{fig_data_quantity}) reaches higher accuracies for smaller training sets as it only has two classes and signal-to-noise for protein signals is higher than for barcodes.

\subsection{ Discussion }

We have shown that convolutional neural networks can accurately classify events from nanopore data. Our network achieves better accuracy than the previous algorithm~\cite{Bell2016a} or manual classification, while at the same time classifying events that were impossible to interpret before. As a result, five times more data can be analysed from the same experiments. Furthermore, the machine learning approach simplifies the analysis by eliminating manual parameter tuning and algorithm development. Instead, we rely on experiments to generate the labels that are used to train the neural network.

In the supplementary material, we use QuipuNet to analyse raw data from other nanopore experiments~\cite{QuipuNet2018}. In~\cite{Kong2017} the authors detected single-nucleotide polymorphisms from the presence of a single binding target. Their designed DNA molecules contain only the sensing region with no barcode. We successfully reproduce results from their analysis using QuipuNet. 
Despite a significantly lower signal-to-noise level for this data set we obtain accuracy of up to 72\% when including folded events. 
If only the unfolded events are analysed, the accuracy is $0.91$. 
This shows that QuipuNet can be readily applied to other nanopore sensing datasets. When designing a nanopore experiment, others should consider the relationship between the desired accuracy and the number of training events.

Our work suggests that deep learning is particularly suitable for nanopore sensing because the experiments can generate large amounts of training data; often with predefined labels. A similar conclusion was reached for nanopore-based DNA sequencing, where a recurrent neural network improves the precision of DNA sequencing~\cite{Boza2017}. Future work may address other difficult problems in the nanopore field. Specifically, peak localisation in noisy datasets~\cite{Kong2016a} can be trained using DNA with known modification positions. Also, running QuipuNet against simulated datasets (generated classically or with generative adversarial networks) could guide the design of the DNA structures in order to maximise the information density or readout accuracy. Both are critically important for information storage on DNA and hold the promise of highly multiplexed protein sensing for medical applications.

\subsection{ Acknowledgements }

We are grateful to Tautvydas Misiunas and Scott Lowe for their advice on the model optimisation. We also thank Nicholas A\@.W\@. Bell and Jinglin Kong for providing  traces from their original experiments. 
K\@.M\@. and U\@.F\@.K\@. acknowledge funding from an ERC consolidator grant (Designerpores 647144). N\@.E.\@ acknowledges funding from the EPSRC, Cambridge Trust and Trinity Hall, Cambridge.


\bibliographystyle{naturemag}
\bibliography{abbreviated}

\end{document}